\documentclass[aps,prc,twocolumn,reprint,floatfix,superscriptaddress,showpacs]{revtex4-1}
\usepackage{graphicx,natbib,amssymb}
\usepackage[pdftex,colorlinks,citecolor=blue,bookmarks]{hyperref}
\usepackage{bm,xcolor,braket,xspace}
\usepackage{amsmath}

\newcommand{\bbz}{$0\nu\beta\beta$\xspace}
\newcommand{\bb}{$\beta\beta$\xspace}
\begin{document}

\title{Testing the importance of collective correlations in neutrinoless $\beta\beta$ decay}

\author{Javier Men\'endez}
\email{menendez@nt.phys.s.u-tokyo.ac.jp}
\affiliation{Department of Physics, University of Tokyo, Hongo, Tokyo 113-0033, Japan}
\affiliation{Institut f\"ur Kernphysik, Technische Universit\"at Darmstadt, D-64289 Darmstadt, Germany}
\affiliation{ExtreMe Matter Institute EMMI, GSI Helmholtzzentrum f\"ur Schwerionenforschung GmbH, D-64291 Darmstadt, Germany}

\author{Nobuo Hinohara}
\email{hinohara@ccs.tsukuba.ac.jp}
\affiliation{Center for Computational Sciences, University of Tsukuba, Tsukuba 305-8577, Japan}
\affiliation{NSCL/FRIB Laboratory, Michigan State University, East Lansing, Michigan, 48824, USA}

\author{Jonathan Engel}
\email{engelj@email.unc.edu}
\affiliation{Department of Physics and Astronomy, University of North Carolina, Chapel Hill, North Carolina, 27599-3255, USA}

\author{Gabriel Mart\'inez-Pinedo}
\email{Gabriel.Martinez@physik.tu-darmstadt.de}
\affiliation{Institut f\"ur Kernphysik, Technische Universit\"at Darmstadt, D-64289 Darmstadt, Germany}
\affiliation{GSI Helmholtzzentrum f\"ur Schwerionenforschung GmbH, D-64291 Darmstadt, Germany}

\author{Tom\'as R. Rodr\'iguez}
\email{tomas.rodriguez@uam.es}
\affiliation{Departamento de F\'isica Te\'orica, Universidad Aut\'onoma de Madrid, E-28049 Madrid, Spain}

\begin{abstract}
We investigate the extent to which theories of collective motion can capture
the physics that determines the nuclear matrix elements governing neutrinoless
double-beta decay.  To that end we calculate the matrix elements for a series
of isotopes in the full $pf$ shell, omitting no spin-orbit partners.  With the
inclusion of isoscalar pairing, a separable collective Hamiltonian that is
derived from the shell model effective interaction reproduces the full
shell-model matrix elements with good accuracy.  A version of the generator
coordinate method that includes the isoscalar pairing amplitude as a coordinate
also reproduces the shell model results well, an encouraging result for
theories of collective motion, which can include more single-particle orbitals
than the shell model.  We briefly examine heavier nuclei relevant for
experimental double-beta decay searches, in which shell-model calculations with
all spin-orbit partners are not feasible; our estimates suggest that isoscalar
pairing also plays a significant role in these nuclei, though one we are less
able to quantify precisely.
\end{abstract}


\maketitle

\section{Introduction}~\label{intro}   

Neutrinoless double-beta (\bbz) decay, if observed, would tell us that
neutrinos are their own antiparticles, i.e.\ Majorana particles.  It has the
potential to reveal the overall neutrino mass scale and hierarchy as well, but
only to the extent that we know the nuclear matrix elements that, together with
the mass scale, determine the decay rate~\cite{avi08}.  The matrix elements,
which must be calculated, are at present quite uncertain~\cite{vog12}, and
reducing that uncertainty is becoming increasingly urgent as decisions on
planning and funding ton-scale $\beta\beta$ decay experiments draw
near~\cite{zub12,cre13,cad15}.

Theorists compute the \bbz-decay nuclear matrix elements in a variety of
models, including the shell model~\cite{men08,nea14}, the interacting boson
model (IBM)~\cite{bar15}, the quasiparticle random-phase approximation
(QRPA)~\cite{mus13,sim13,hyv15} and the generator coordinate method
(GCM)~\cite{vaq13,yao14,hin14}, the last two of which can incorporate energy
density functional (EDF) theory~\cite{mus13,vaq13,yao14}.  The GCM and the IBM
pay particular attention to collective phenomena such as deformation and
pairing, and neglect, for the most part, non-collective correlations.  The
prominence of these collective approaches makes it important to know how
accurate they can be or, in other words, the extent to which collective
correlations determine the \bbz decay nuclear matrix elements.  And among all
possible collective correlations, which are the most relevant for \bbz decay?
These questions have never been addressed in a systematic way.

Although we will not provide conclusive answers, we can fully address the
questions in nuclei for which shell model calculations in a full
harmonic-oscillator shell are feasible.  A full-shell valence space is useful
because it guarantees spin-orbit partners for all orbitals.  Spin-orbit pairs
are important because the two-body matrix elements of the \bbz decay operator between
these orbitals are typically large.  Spin-orbit partners are needed,
furthermore, to fully capture the effects of isoscalar spin-one pairing between
protons and neutrons.  The $pf$ shell is a prime example of a space in which
each orbital has a spin-orbit partner, and shell model calculations there are
very successful~\cite{cau05}.  They can describe collective phenomena such as
deformation despite neglecting cross-shell correlations, which are not
very important for ground-state properties of lower $pf$-shell nuclei close to
stability~\cite{cau05}.

We thus explore the role of collective correlations in the $pf$ shell, testing
to see which collective degrees of freedom have the largest effect on \bbz
decay matrix elements and the degree to which the most important degrees of
freedom determine the matrix elements. In larger valence spaces, which at
present are still beyond what the shell model can treat well, the role of
collective correlations is only expected to be greater, at least away from
shell closures.  Of course in most $pf$-shell nuclei $\beta\beta$ decay is
either energetically forbidden or exceedingly slow compared to single-$\beta$
decay; nevertheless, nuclear matrix elements can be calculated and their
quality assessed.  Some recent papers \cite{rod13,men14} have taken a similar
approach, gaining insight into $\beta\beta$ decay through the systematic
calculation of matrix elements in nuclei that could not themselves be used in
$\beta\beta$ decay experiments.

Here, we conduct two kinds of tests.
First, we extract from Ref.\ \cite{duf96} the separable collective Hamiltonian
that best approximates a full shell-model effective interaction in the $pf$
shell.  This Hamiltonian employs a monopole interaction and collective pieces:
isovector $J=0$ and isoscalar $J=1$ pairing terms, a quadrupole-quadrupole
term, and a spin-isospin term.  We compare the \bbz decay matrix elements that
this interaction produces with those produced by the full shell model
interaction in the Ca, Ti and Cr isotopic chains (heavier elements are
computationally more demanding, as well as more sensitive to orbitals beyond
the $pf$ shell), and identify the most relevant collective correlations for
$\beta\beta$ decay.  Second, we use the collective interaction within a GCM
calculation that includes the isoscalar pairing amplitude and the quadrupole
moment as generator coordinates, and compare the resulting \bbz decay matrix
elements to those of the shell model.  Finally, we try to assess the degree to
which our conclusions hold for the heavier nuclei in which $\beta\beta$ decay
could be detected in next-generation experiments.

The rest of this paper is structured as follows: Section~\ref{interaction}
describes the extraction of the separable collective interaction and discusses
each of its components.  Section \ref{results} briefly presents the \bbz decay
operator and compares the matrix elements, calculated in the shell model with
both the full and collective Hamiltonians, for isotopes of Ca, Ti and Cr.  It
also shows GCM matrix elements for the same nuclei, calculated with the same
collective interaction, and finally discusses the matrix elements for heavier
nuclei that are of real interest for \bbz decay experiments.  Section
\ref{summary} is a conclusion.

\section{Separable collective interaction}~\label{interaction}

We work in the $pf$-shell configuration space, comprising the $0f_{7/2}$,
$1p_{3/2}$, $1p_{1/2}$ and $0f_{5/2}$ orbitals.  As a reference Hamiltonian we
use the shell model interaction KB3G~\cite{pov01}, which has been extensively
tested throughout the $pf$-shell.  This interaction provides a very good
description of nuclear structure, including spectroscopy, electromagnetic and
Gamow-Teller transitions, and deformation~\cite{cau05}.  Then, following the
work of Dufour and Zuker~\cite{duf96}, we build the separable collective
Hamiltonian that best approximates KB3G.  Roughly speaking, Ref.~\cite{duf96}
determines the structure of the lowest-lying collective states in the
particle-hole and pairing representations with a given angular momentum $J$,
isospin $T$, and parity $\pi$, and then constructs a series of separable terms,
with appropriate strengths, that reproduce those states.  Dufour and Zuker find
that the most important terms in the particle-hole channel are the isoscalar
quadrupole and spin-isospin ($\sigma \tau \sigma \tau$) interactions, and in
the pairing channel the isovector $J^\pi=0^+$ and isoscalar $J^\pi=1^+$
interactions.

\begin{table}[t]
\caption{Strengths (in MeV) of the isovector pairing ($g^{T=1}$), isoscalar
paring ($g^{T=0}$), spin-isospin ($g_{ph}$), and quadrupole ($\chi$)
interactions in the separable collective Hamiltoninan $H_\text{coll}$
[Eq.~(\ref{eq:hsep})].  The values are taken from Ref.~\cite{duf96} and scaled
to nucleon number $A=42$.  For heavier isotopes the strengths are multiplied by
$(42/A)^{1/3}$.}
\begin{center}
\begin{ruledtabular}
\begin{tabular*}{0.6\columnwidth}{@{\extracolsep{\fill}}cccc}
$g^{T=1}$ & $g^{T=0}$ & $g_{ph}$ & $\chi$ \\ \hline
$-0.377$ & $-0.587$ & $0.057$ & $-0.141$ \\
\end{tabular*}
\end{ruledtabular}
\end{center}
\label{tab:couplings}
\end{table}

The separable collective Hamiltonian, $H_\text{coll}$, that includes the full
monopole piece of the KB3G interaction and the dominant collective terms found
by Dufour and Zuker has the form
\begin{align}
\label{eq:hsep}
H_\text{coll} &= H_M + g^{T=1} \sum_{n=-1}^1 {S}_n^\dag {S}_n   + g^{T=0} \sum_{m=-1}^1
{P}_m^\dag {P}_m  \nonumber \\ 
&+g_{ph} \sum_{m,n=-1}^1 : {\mathcal{F}}^\dag_{mn} {\mathcal{F}}_{mn} :
+ {\chi} \sum_{\mu=-2}^2 : Q_\mu^\dag Q_\mu : \,, 
\end{align}
where the colons indicate normal ordering.  The monopole Hamiltonian $H_M$
includes two-body terms and one-body (single-particle) energies, both taken from
KB3G.  In addition
\begin{align}
\label{eq:ops-def}
S^\dag_n &= \frac{1}{\sqrt{2}} \sum_{\alpha} \sqrt{2l_\alpha+1} 
\left( a_\alpha^\dag a_\alpha^\dag \right)^{0,0,1}_{0,0,n}\,, \nonumber \\
P^\dag_m &= \frac{1}{\sqrt{2}} \sum_{\alpha} \sqrt{2l_\alpha+1} 
\left( a_\alpha^\dag a_\alpha^\dag \right)^{0,1,0}_{0,m,0}\,, \nonumber \\
{\cal F}_{mn} &= 2 \sum_{\alpha} \sqrt{2l_\alpha+1} \left( a_\alpha^\dag
\tilde{a}_\alpha \right)^{0,1,1}_{0,m,n}\,, \\
Q_\mu &= \frac{1}{\sqrt{5}} \sum_{\alpha,\beta}\bra{n_\alpha l_\alpha}\!|r^2Y_2/b^2
|\!\ket{n_\beta l_\beta} \left(a_\alpha^\dag \tilde{a}_\beta
\right)^{2,0,0}_{\mu,0,0} \,, \nonumber 
\end{align} 
where $\mathcal{F}_{mn}$, written in first quantization, is just $\sum_i
\sigma_m(i) \tau_n(i)$, $b$ is the usual oscillator parameter, $a^\dag_\alpha$
creates a nucleon in a single-particle orbital with principal quantum number
$n_\alpha$ and orbital angular momentum $l_\alpha$, and $\tilde{a}_{a}$ destroys a
nucleon in the time-reversed orbital (more precisely,
$\tilde{a}_{l_\alpha,m_\alpha,s_\alpha,\tau_\alpha} \equiv (-1)^{l_\alpha+1
-m_\alpha-s_\alpha-\tau_\alpha} a_{l_\alpha,-m_\alpha,s_\alpha,-\tau_\alpha}$,
where $m_\alpha$ is the $z$ component of the orbital angular momentum,
$s_\alpha$ is the $z$-component of the spin, and $\tau_\alpha$ is the
$z$-component of the isospin). The superscripts following the parentheses stand
for the two-particle orbital angular momentum, spin, and isospin, and the
subscripts for their $z$-components.  The strengths of the various terms,
$g^{T=1}$, $g^{T=0}$, $g_{ph}$ and $\chi$, are taken from Ref.~\cite{duf96} and
appear in Table~\ref{tab:couplings} for mass $A=42$ (they scale with
$A^{-1/3}$).  Note that the pairing and quadrupole-quadrupole terms are
attractive, as expected.  Reference~\cite{mar99} uses a similar collective
Hamiltonian, also based on the decomposition in Ref.~\cite{duf96}, but without
the spin-isospin term, to study the competition between isovector and isoscalar
pairing in $pf$-shell nuclei.

The significance of the various terms in $H_\text{coll}$ is as follows: The
monopole Hamiltonian $H_M$ adds effective neutron- and proton-number-dependent
effective single-particle energies to the bare energies.  The remaining terms
are collective --- an isovector spin-$0$ pairing interaction, an isoscalar
spin-$1$ pairing interaction, a quadrupole-quadrupole interaction,
and a Landau-Migdal-style spin-isospin interaction. Many studies of nuclear collectivity
(e.g.\ \cite{RevModPhys.35.853,bar65,bar68}) include only isovector pairing
(usually without the proton-neutron part) and quadrupole-quadrupole terms.
And isoscalar pairing is frequently downplayed.
Among the models studying \bbz decay matrix elements, the EDF-based GCM and the
IBM have not yet included isoscalar pairing explicitly.

\begin{figure}[t]
\includegraphics[width=\columnwidth]{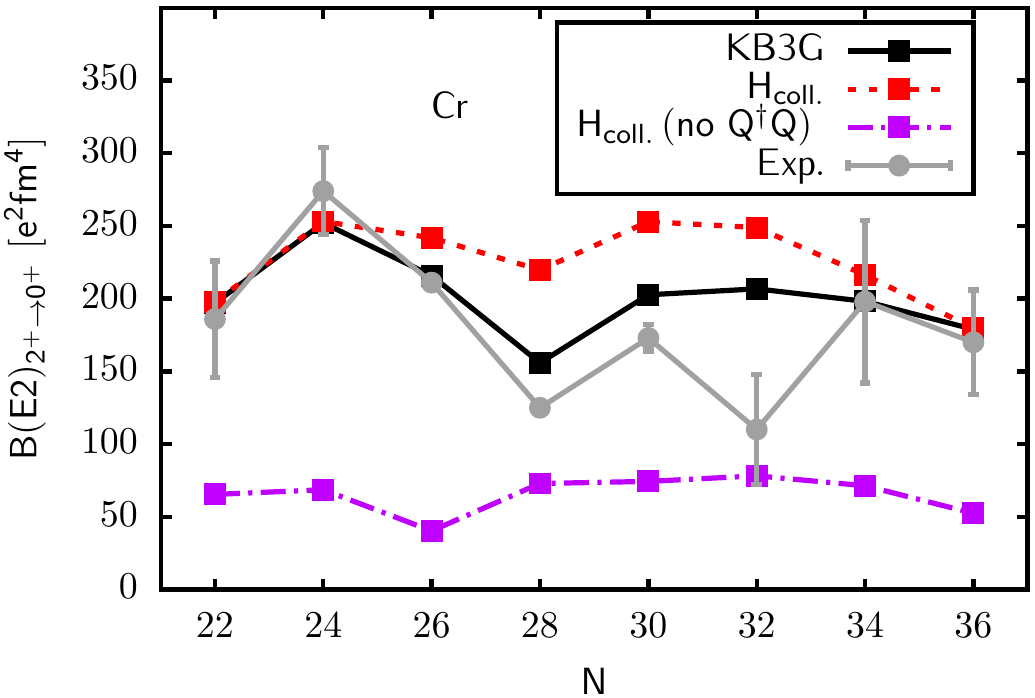}
\caption{\label{fig:be2}$B(E2)$ values for transitions from the
first $2^+$ state to the ground state in Cr isotopes, as a function of mass
number $A$.  Experiment~\cite{adndt} (gray circles with error bars, connected by
solid line) is compared to calculations with several interactions.  Black
squares (solid line) show results produced by the KB3G interaction, red squares
(dashed line) by the full collective interaction $H_\text{coll}$, and purple
squares (dotted line) by $H_\text{coll}$ without the quadrupole-quadrupole part.
Protons have an effective charge of 1.5$e$ and neutrons of 0.5$e$.} 
\end{figure}

According to Ref.\ \cite{duf96}, the terms included in $H_\text{coll}$ are the
most important for $pf$-shell nuclei (we could also have included, for example,
an isovector quadrupole-quadrupole interaction, a hexadecupole-hexadecupole
piece, or an isovector $J=2$ pairing term).  We can test the adequacy of
$H_\text{coll}$ by comparing its results to those of the KB3G interaction.
Figure\ \ref{fig:be2} shows calculated and experimental $B(E2)$ strengths from
the lowest $2^+$ state to the ground state in even-mass Cr isotopes, which have
enough valence-space nucleons to exhibit collective behavior.  The strengths
produced by $H_\text{coll}$ and KB3G agree nicely, both with each other and experiment
(however KB3G describes the data better).  A similar picture emerges from
a comparison of $B(E2)$ strengths for Ca, Ti and Fe isotopes.  The agreement
suggests that $H_\text{coll}$ captures the correlations necessary for a good
description of low-energy states in the $pf$-shell nuclei.

One advantage of the separable Hamiltonian is that we can selectively include
or exclude particular terms to test their importance for \bbz decay or for any
other observable.  It is well known that individual terms play unique roles in
nuclear structure spectroscopy.  The quadrupole-quadrupole interaction, for
instance, is crucial for generating deformation and shape vibrations and thus
for a good description of $B(E2)$ strengths.  Fig.\ \ref{fig:be2} illustrates
this fact; in the Cr isotopes the transition strengths are drastically
suppressed when the quadrupole-quadrupole term is removed from the
interaction.  By contrast, the removal of other terms, such as isoscalar
pairing, has only a small effect on the $B(E2)$'s.

\section{Results for double-beta decay}~\label{results}

Here we examine the importance of the various parts of the separable collective
Hamiltonian for \bbz decay and the usefulness of models based on a collective
picture, such as the GCM, for calculating the associated nuclear matrix
elements.  These can be written in the form 
\begin{equation}
M_{0\nu}=M_{0\nu}^{GT}-\left(\frac{g_{V}}{g_{A}}\right)^{2}M_{0\nu}^{F}+M_{0\nu}^{T},
\label{eq:NME}
\end{equation}
with $g_{V}=1.0$ and $g_A=1.27$ the vector and axial coupling constants,
respectively.  A detailed definition of each term above appears in Ref.\
\cite{sim99}.  The Gamow-Teller (GT) matrix element, $M_{0\nu}^{GT}$, is the
largest; in the nuclei studied here it accounts for about $85\%$ of the total
matrix element and for the sake of convenience we omit the other terms from the 
discussion to come.
In addition, we use the closure approximation (with closure
energy parameter $\langle E\rangle=7.72$ MeV), which introduces an error of at
most 10\% in both shell model and QRPA calculations~\cite{pan90,sen13}.  We
also ignore the effects of two-body weak currents, studied in
Refs.~\cite{men11,eng14}, and assume simple Argonne-type short range
correlations \cite{sim09}.  None of these choices and approximations affect any
of our conclusions.  Finally, we restrict our study to isotopic chains in the
lower part of the $pf$-shell, where exact shell-model calculations are possible
and higher-lying single-particle orbitals are not relevant for ground-state properties.

\subsection{\bbz decay of $pf$-shell nuclei in the shell model}~\label{sm_results}

First we turn to the ability of the collective interaction $H_\text{coll}$ to
reproduce the shell model GT matrix elements in Eq.\ (\ref{eq:NME}) that are
obtained with the full KB3G shell model interaction \cite{cau08,men08}.  In addition, to explore
the roles of the various collective pieces in $H_\text{coll}$, we perform
additional calculations, each time removing some combination of the
isovector/isoscalar pairing, quadrupole-quadrupole and spin-isospin terms.  We
never remove the monopole part, however, because its role is simply to fix the
energies of single-particle orbitals. 

\begin{figure}
\includegraphics[width=\columnwidth]{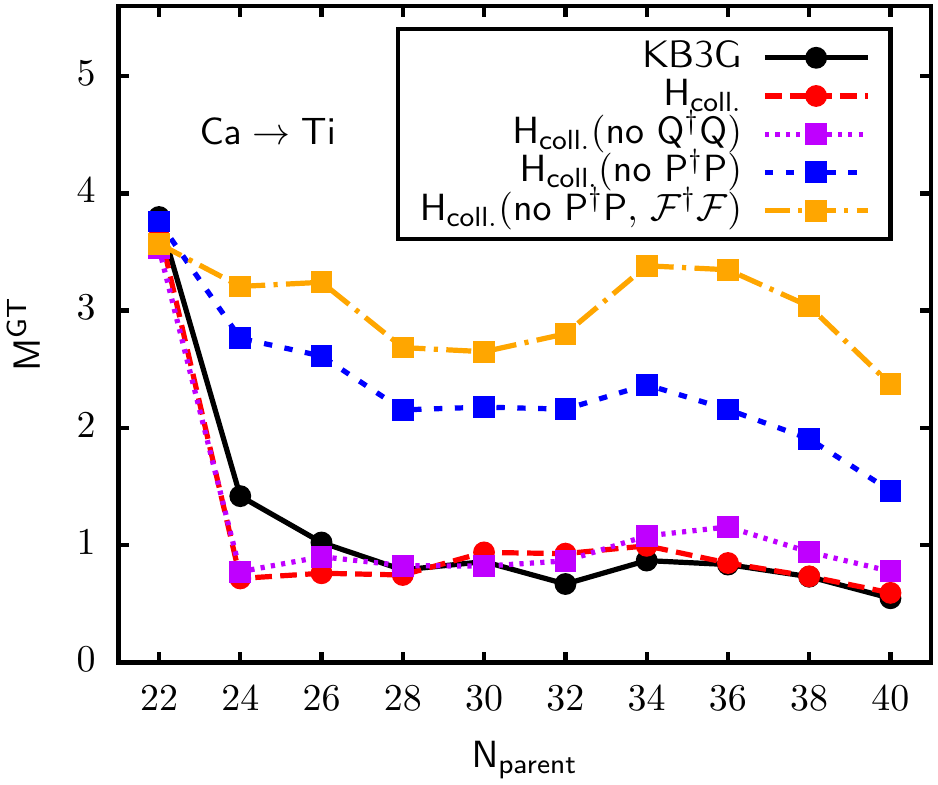}
\caption{Gamow-Teller part of the \bbz decay matrix elements
$M_{0\nu}^{GT}$, for the decay of Ca isotopes into Ti as a function of the
neutron number $N_\text{parent}$ in the parent nucleus.  Results are shown for
the KB3G interaction (black circles, solid line), the full collective
interaction $H_\text{coll}$ (red circles, dashed line), $H_\text{coll}$ with the
quadrupole-quadrupole term removed (purple squares, dotted line),
$H_\text{coll}$ with the isoscalar pairing term removed (blue squares,
short-dashed line), and $H_\text{coll}$ with both the isoscalar-pairing and
spin-isospin pieces removed (orange squares, dot-dashed line).
}\label{fig:ca_pqq} 
\end{figure}

Figure~\ref{fig:ca_pqq} shows that the matrix elements obtained from the full
$H_\text{coll}$ are very close to those obtained from the KB3G interaction.
That a simple separable collective Hamiltonian can capture the main features of
the matrix element suggests that fine details of the nuclear interaction affect
$0\nu\beta\beta$ decay only moderately.  The result confirms and significantly
extends the findings of Refs.\ \cite{hor10,men09}, which contained relatively
small variations --- between 10\% and 20\% --- in the matrix elements obtained
from different shell-model effective interactions.

Figure~\ref{fig:ca_pqq} also shows that the matrix elements change relatively
little when the quadrupole-quadrupole interaction is excluded. This result holds
not only for the mostly spherical Ca and Ti isotopes involved in the decay
matrix elements of Ca, but also for the Ti and Cr isotopes (not shown), which
involve deformed nuclei, as Fig.~\ref{fig:be2} indicates.  This apparent lack of
sensitivity to quadrupole correlations contrasts with the results of several
studies in the shell model, EDF-based GCM, and QRPA that point to a reduction of
the \bbz decay matrix elements between deformed nuclei~\cite{men10,rod10,fan11,mus13}. 
The reason is probably the moderate deformation of the $pf$-shell nuclei considered here.  
If the quadrupole
correlations are increased by doubling the strength of the quadrupole-quadrupole interaction, 
the corresponding matrix elements shrink, as expected, by $30\%-40\%$.
In addition, when the quadrupole-quadrupole interaction is excluded 
from the calculation of the parent nuclei (but not the daughter nuclei), the matrix elements decrease by $15\%-20\%$.  That result is consistent with those of previous studies ~\cite{men10,rod10,fan11,mus13} that note a small matrix element when the parent and daughter have different quadrupole properties.

\begin{figure}
\includegraphics[width=\columnwidth]{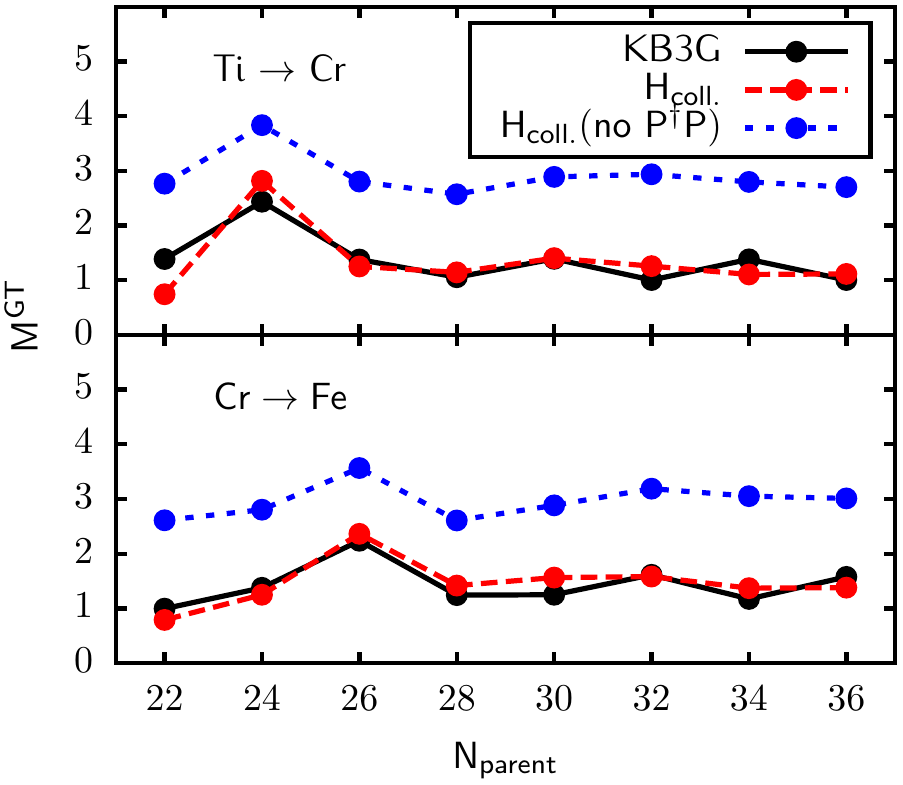}
\caption{Gamow-Teller part of the \bbz decay matrix elements,
$M_{0\nu}^{GT}$, for the decay of Ti isotopes into Cr (top panel), and Cr
isotopes into Fe (bottom), as a function of the neutron number $N_\text{parent}
$ of the parent nucleus.  Results are shown for the KB3G interaction (black,
solid line), the collective interaction $H_\text{coll}$ (red, dashed line), and
$H_\text{coll}$ without the isoscalar pairing term (blue, short-dashed line).
}\label{fig:ca_st} 
\end{figure}

Perhaps the most striking feature of Fig.\ \ref{fig:ca_pqq} is the suppression
of the matrix elements by isoscalar pairing.  Removing that term from the
Hamiltonian increases the matrix elements by more than a factor of two (closer
to three in many isotopes), or between 1 and 2 units.
When, in addition, the
spin-isospin term is removed, the matrix elements grow even further.  As Fig.\
\ref{fig:ca_st} shows, the large effect of isoscalar pairing is common to the
matrix elements of all the Ca, Ti, and Cr isotopes we study, from those with
$N\sim Z$ to very neutron-rich nuclei.  For the matrix elements of the most
isospin-asymmetric nuclei ($^{58}$Ca and $^{60}$Ca) the effect of isoscalar
pairing is somewhat milder but still important.  The sensitivity to isoscalar
(proton-neutron) pairing is familiar from QRPA~\cite{vog86,eng88} and GCM
studies~\cite{hin14} and makes it clear that a good description of
proton-neutron correlations is crucial to obtain accurate \bbz decay nuclear
matrix elements.

The significance of isoscalar pairing is not quite as straightforward as it
first appears, however.
The matrix elements vary just about 10\% when only the spin-isospin interaction
is omitted from $H_\text{coll}$.
As Fig.\ \ref{fig:ca_pqq} shows, when the spin-isospin
term is included in the separable collective Hamiltonian, the impact of
omitting isoscalar pairing, though still significant, is smaller than with the
spin-isospin term excluded.
This result suggests that the missing isoscalar-pairing correlations can to
some extent be compensated for, or captured, by other collective interactions.
In that sense, we can consider the dramatic changes in the matrix elements
shown in Figs.~\ref{fig:ca_pqq} and \ref{fig:ca_st} to be an upper bound for
the effects of isoscalar pairing.  Pieces of the nuclear Hamiltonian, both
collective and non-collective, that are not included in $H_\text{coll}$ might
soften the impact of omitting isoscalar-pairing, in the same way that the
spin-isospin interaction does.

\begin{figure}
\includegraphics[width=\columnwidth]{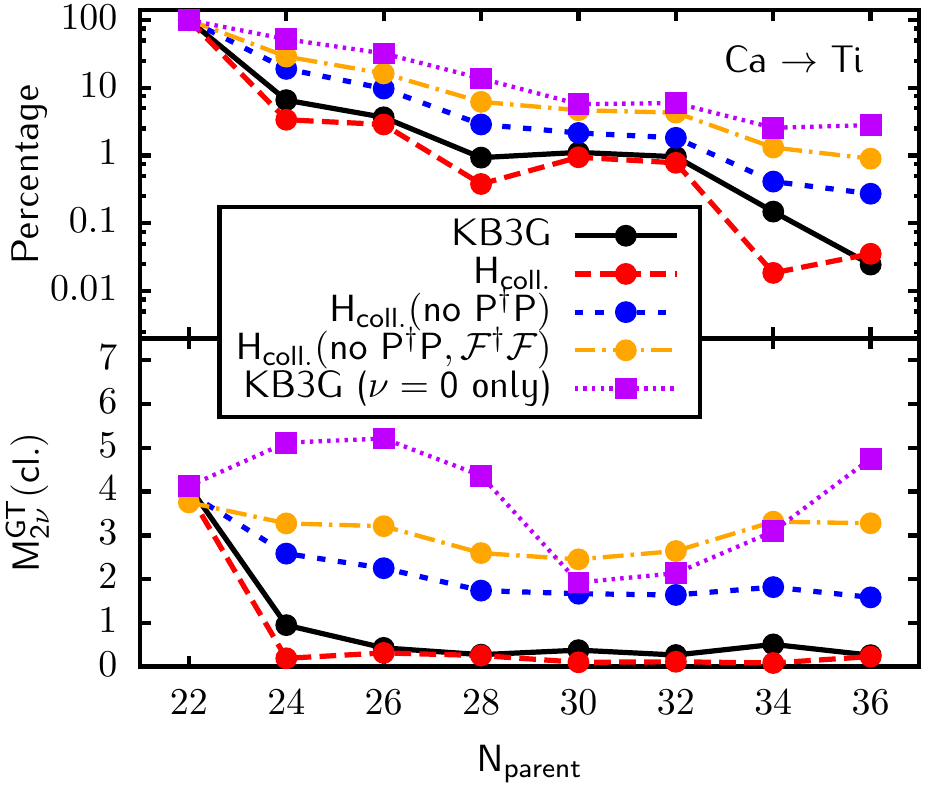}
\caption{\textbf{Top panel:} Percentage of ground state in
daughter nuclei (Ti isotopes) belonging to $SU(4)$ irreducible representations
(irreps) that are also present in the corresponding parent nuclei (Ca isotopes),
as a function of the neutron number $N_\text{parent}$ of the parent nucleus.
Results are shown for the KB3G effective interaction (black circles, solid
line), the collective Hamiltonian $H_\text{coll}$ (red circles, dashed line),
$H_\text{coll}$ without the isoscalar pairing term (blue circles, short-dashed
line), $H_\text{coll}$ without both the isoscalar pairing and the spin-isospin
terms (orange circles, dot-dashed line), and the KB3G interaction diagonalized
in a basis of a seniority-zero states (purple squares, dotted line).\\ 
\textbf{Bottom Panel:} $M^{GT}_{2\nu}(\text{cl.})$ (see text) as a function of
$N_\text{parent}$.  Correspondence between results and symbols/lines is the
same as in the top panel.}
\label{fig:irreps} 
\end{figure}

The impact of isoscalar pairing correlations in \bbz decay is
undeniable.  One way to understand it is through spin-isospin $SU(4)$
symmetry.  The GT operator, if we neglect the neutrino potential, is invariant
under $SU(4)$ transformations, implying that only states belonging to the same
irreducible representations (irreps) of $SU(4)$ can be connected by the
operator; the matrix elements between states in different irreps vanish.
Furthermore, in the absence of spin-orbit splitting in the $H_M$ piece, the
collective Hamiltonian $H_\text{coll}$ is invariant under $SU(4)$ if the
isovector and isoscalar pairing terms have the same strength,
$g^{T=1}=g^{T=0}$.  The situation resembles that associated with the \bb decay
Fermi operator, which because of isospin symmetry has vanishing matrix elements
between states belonging to different isospin-$SU(2)$ irreps, i.e.\ having
different total isospin~\cite{pri59}.  In $0\nu\beta\beta$ decay the neutrino
potential breaks the $SU(2)$ invariance of the operator and the matrix
elements, $M_{0\nu}^F$, do not vanish, but they are nevertheless suppressed
~\cite{men08,men14,bar15,sim13}.

In $pf$-shell nuclei the spin-orbit splitting is sizable, and nuclear states
are in general a combination of several different $SU(4)$ irreps~\cite{vog93}.
However, since $g^{T=0}$ is only slightly larger than $g^{T=1}$, and the
spin-isospin interaction, which conserves the $SU(4)$ symmetry, effectively
increases the energy separation among $SU(4)$ irreps, the fraction of irreps
shared between the parent and daughter nuclei is small.  This fact is
illustrated in the top part of Fig.~\ref{fig:irreps}, which shows the
percentage of the ground state in each Ti isotope (daughter nucleus) belonging
in irreps that are also present in the ground state of the corresponding Ca
isotope (parent).  The small percentages mean that in the approximation that
the neutrino potential is replaced by a constant, i.e.\ with the \bbz decay
operator replaced by the closure version of the $2\nu\beta\beta$ decay operator
$M^{GT}_{2\nu}(\text{cl.)}$, the matrix elements are tiny (see the bottom panel
of Fig.~\ref{fig:irreps}).  The result explains why $M^{GT}$ for \bbz decay,
which reflects mild $SU(4)$-breaking by the neutrino potential, is generally
small rather than either tiny or large.  The only exception is in mirror
nuclei, where the irreps in the parent and daughter are identical.  There the
matrix elements are larger than others in the same isotopic chain, as both
shell-model and GCM calculations show~\cite{men08,men14,rod13}.

Little remains of $SU(4)$ symmetry when the isoscalar pairing and the
spin-isospin terms are removed from the Hamiltonian.  As Fig.~\ref{fig:irreps}
shows, setting $g^{T=0}=0$ causes the percentage of the ground states in parent
and daughter nuclei belonging to shared $SU(4)$ irreps to increase substantially, which in turn
increases $M^{GT}_{2\nu}(\text{cl.})$.  The effect is even stronger when the
spin-isospin interaction is removed as well.  And as Figs.~\ref{fig:ca_pqq}
and~\ref{fig:ca_st} show, the $M^{GT}_{0\nu}$ matrix elements also increase
dramatically.
(The  percentage of common irreps in the parent and and daughter nuclei
actually decreases faster with $N-Z$ than the matrix elements.  The reason is  that the matrix 
elements between states in the same irrep are proportional to  $N-Z$~\cite{vog88}.)

The same kind of $SU(4)$ breaking is at play when ground states
are forced to have seniority zero, that is, states consisting entirely of
like-particle $J=0$ pairs.  By construction, seniority-zero states have no
proton-neutron pairs or spin-isospin correlations and thus break $SU(4)$
strongly.  As a result, the percentage of the ground states in the parent and
daughter nuclei belonging to shared irrep increases and both
$M^{GT}_{2\nu}(\text{cl.})$ and $M^{GT}_{0\nu}$ grow (see Refs.\
\cite{cau08,men14}).

\begin{figure}
\includegraphics[width=\columnwidth]{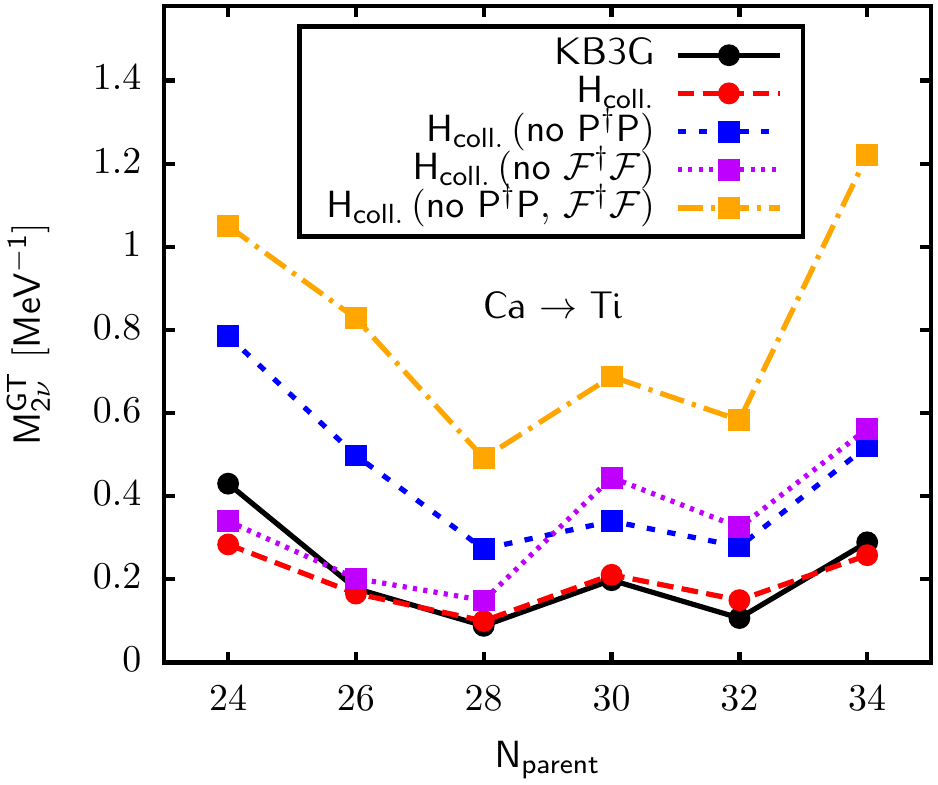}
\caption{Two-neutrino \bb decay matrix elements
$M_{2\nu}^{GT}$, for the decay of Ca isotopes into Ti as a function of the
neutron number $N_\text{parent}$ in the parent nucleus.  Results are shown for
the KB3G interaction (black circles, solid line), the full collective
interaction $H_\text{coll}$ (red circles, dashed line),
$H_\text{coll}$ with the isoscalar pairing term removed (blue squares,
short-dashed line), and $H_\text{coll}$ with both isoscalar-pairing and
spin-isospin parts removed (orange squares, dot-dashed line).
}\label{fig:2nu} 
\end{figure}

In addition, we study the impact of isoscalar pairing in $2\nu\beta\beta$ decay.
The lower part of Fig.~\ref{fig:irreps} suggests that spin-isospin and 
isoscalar pairing correlations are relevant for $2\nu\beta\beta$ decay,
but for a detailed study the matrix elements need to be calculated beyond
the closure approximation, because of the small momentum transfers involved in
 $2\nu\beta\beta$ transitions~\cite{vog12}.
 
Figure~\ref{fig:2nu} shows
non-closure $2\nu\beta\beta$ decay matrix elements calculated with the shell-model
KB3G interaction, the collective Hamiltonian $H_{\text{coll}}$,
and with the same Hamiltonian but excluding the isoscalar pairing
and/or spin-isospin parts in $H_{\text{coll}}$.
As in $0\nu\beta\beta$ decay, the results obtained with 
the full collective Hamiltonian $H_{\text{coll}}$
are in very good agreement with the full shell-model results,
suggesting that the collective Hamiltonian includes all the interaction components
relevant for $2\nu\beta\beta$ decay and that fine details
of the shell-model interaction only affect this decay moderately.

The impact of isoscalar pairing and spin-isospin correlations is sizable and, 
like in $0\nu\beta\beta$ decay, excluding both collective terms
(or only the isoscalar pairing part)
leads to significantly overestimated $2\nu\beta\beta$ decay matrix elements. 
Figure~\ref{fig:2nu} also shows that for $2\nu\beta\beta$ decay,
excluding only the spin-isospin interaction leads to overestimated matrix elements 
in neutron-rich nuclei as well
(in $0\nu\beta\beta$ decay all matrix elements vary just by about 10\%).
In general, the effect of excluding both isoscalar pairing and 
spin-isospin terms is larger than the sum of the matrix element 
increases resulting from not including each term individually.
Overall the impact of the isoscalar pairing and spin-isospin terms
are qualitatively similar but quantitatively different
in the neutrinoless and two-neutrino \bb decay modes.

\subsection{\bbz decay of $pf$-shell nuclei in the GCM}~\label{gcm_results}

The strength of the GCM, QRPA (based on EDF) and IBM (based on bosons) is their
treatment of collectivity.  Although these methods sacrifice some of the
complex valence-space correlations captured by the shell model, they can
effectively include larger single-particle spaces, which are frequently
required to capture collective correlations.  
Here we test the ability of the GCM, with the same collective interaction
discussed in Sec.~\ref{interaction}, $H_{\text{coll}}$, to reproduce
shell-model $M^{GT}_{0\nu}$ matrix elements. 

The GCM is an extension of mean-field theory that supplements the lowest-energy
quasiparticle vacuum with other quasiparticle vacua that are constrained to
have different expectation values for the operators representing collective
coordinates.  The method is used most commonly to allow vacua with a range of
values for the axial quadrupole moment $\braket{Q_{0}}$ to appear in low-lying
collective states; in such applications the quantum states are obtained by
diagonalizing the Hamiltonian in the space of non-orthogonal vacua with
different quadrupole moments, or equivalently, different values of the deformation parameter
$\beta$.

The generator coordinates, the collective degrees of freedom in the GCM, are
chosen at the beginning of the calculation, and it is crucial to include all collective degrees of
freedom that are important for the phenomena being studied.  Non-axial
quadrupole coordinates are ostensibly important but because they affect ground
states less than excited states (and because they make angular-momentum
projection quite time consuming), we restrict ourselves to axially deformed
shapes.  And we neglect like-particle pairing fluctuations because they change  
$M_{0\nu}^{GT}$ by 30\% or less in the $pf$ shell~\cite{vaq13}.
Isoscalar pairing is another story, however.  Beginning with the QRPA work of
Ref.\ \cite{eng88}, it has been apparent that dynamical isoscalar pairing
correlations have a significant effect on $M_{0\nu}^{GT}$ (the static
correlations vanish).  Reference\ \cite{hin14} showed how to add their effects
by using $\braket{P_0+P_0^{\dag}}$ as a generator coordinate.  This isoscalar
pairing amplitude breaks the particle number and rotational symmetries but
preserves axial symmetry, so we project the HFB states onto states with good
particle number and angular momentum to restore the broken symmetries. The
other components of the isoscalar pairing amplitude (related to $P^\dag_{\pm
1}$) are included through the angular-momentum projection. Isospin symmetry is
broken and not restored in our calculation.

\begin{figure}[t]
\includegraphics[width=\columnwidth]{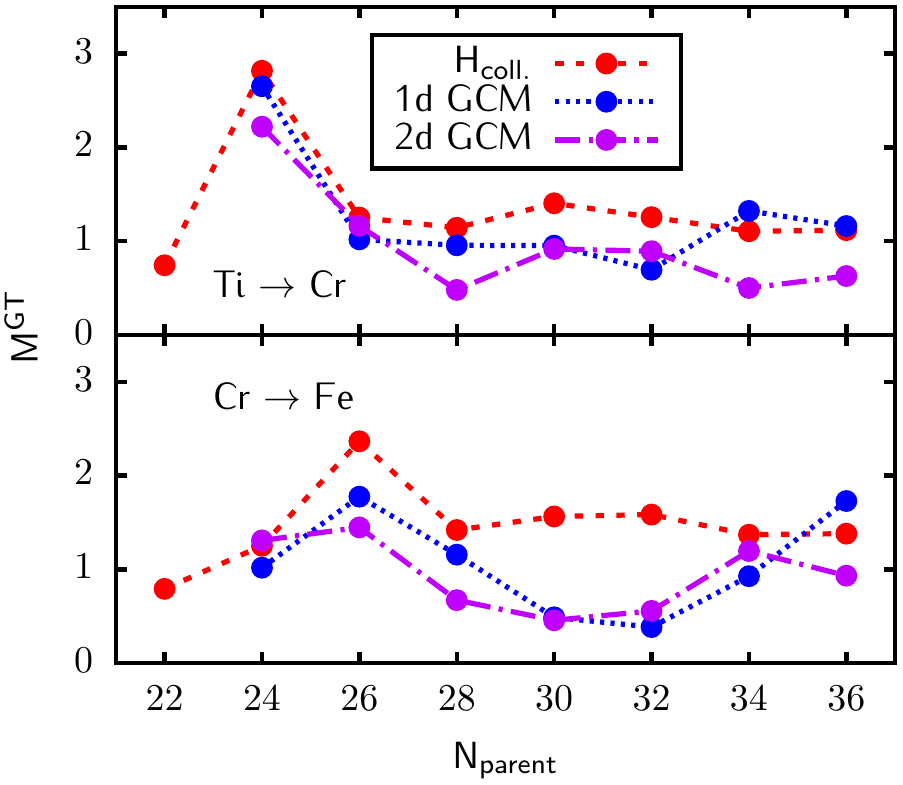}
\caption{Gamow-Teller part of \bbz decay matrix elements,
$M^{GT}_{0\nu}$, for the decay of Ti isotopes into Cr (top panel) and Cr
isotopes into Fe (bottom), as a function of the neutron number $N_\text{parent}$
in the parent nucleus.  Results are shown for the shell model with the
collective Hamiltonian $H_\text{coll}$ (red, dashed line), the GCM with the same $H_\text{coll}$ but without quadrupole-quadrupole 
interaction,
and with the isoscalar pairing amplitude as only coordinate (blue,
short-dashed line), and the GCM with the quadrupole-quadrupole interaction and
with the axial quadrupole deformation parameter $\beta$ as second coordinate
(purple, dotted line).}
\label{fig:gcm} 
\end{figure}

Figure \ref{fig:gcm} shows our GCM results for the \bbz decay of Ti and Cr
isotopes with the interaction $H_{\text{coll}}$.
We compare them to the
shell-model values obtained from the diagonalization of  $H_{\text{coll}}$.  Both the ``1d'' 
version of the GCM, which treats only the isoscalar pairing amplitude as a coordinate, and the 
``2d"
version, which adds a coordinate corresponding to axial deformation, agree well
with the full shell-model results with $H_\text{coll}$.  (The two GCMs agree with each 
other because the addition of the quadrupole interaction,
as we have already seen, does not have a large
effect on the matrix elements.)  Together with
the demonstrated adequacy of $H_\text{coll}$, the agreement suggests that
theories of collective motion, which can be extended to several shells, can
provide reliable matrix elements in heavier nuclei, where a single valence
shell may not be sufficient.

In the isotopes with neutron numbers in the range $N=28-32$ the GCM
results deviate from those of the shell-model.  For these transitions either
the parent or daughter nucleus contains a closed shell at $N=28$ or $N=32$,
and collectivity plays a smaller role.  In addition, at present our GCM
calculation excludes vacua without pairing to avoid numerical instability, so that we
omit the most important states in closed-shell systems.  
The inclusion of individual particle-hole excitations across shells in the GCM
basis will improve the present results.

\subsection{\bbz decay in important nuclei near $A=80$ and $A=130$}~\label{heavy}

The results presented so far illustrate the importance of collective
correlations for the $0\nu\beta\beta$ decay matrix elements of nuclei in the
lower part of the $pf$-shell.  Of all these isotopes, however, only $^{48}$Ca
actually has even a chance to be used in a \bb experiment.  All other relevant
nuclei are too heavy for shell-model calculations in complete oscillator shells,
so that an analysis like that in Sec.\ \ref{sm_results} is not possible.
Nevertheless, we try to estimate the importance of isoscalar pairing for the
\bb decay of these isotopes.

We consider two different valence spaces in calculating the matrix elements for
the $0\nu\beta\beta$ decay candidates heavier than $^{48}$Ca.  In $^{76}$Ge and
$^{82}$Se, the valence space comprises the $1p_{3/2}$, $1p_{1/2}$, $0f_{5/2}$
and $0g_{9/2}$ orbitals, while for $^{124}$Sn, $^{130}$Te, and $^{136}$Xe, it
comprises the $1d_{5/2}$, $2s_{1/2}$, $0g_{7/2}$, $0d_{3/2}$ and $0h_{11/2}$
orbitals.  In each case, two spin-orbit partners are missing from the
shell-model space.  We base our calculations on the shell model GCN2850
effective interaction in the first space and the GCN5082 interaction in the
second \cite{men08}.

One drawback of incomplete oscillator shells and missing spin-orbit partners is
that isoscalar pairing is inhibited.  The single-particle orbitals included in
the shell model calculations are bounded by large gaps, so that the inhibition
is not artificial, but the ability of pairing correlations to transcend the
shell gaps cannot be fully evaluated without explicitly including all spin-orbit
partners.  In addition, it is not appropriate in such valence spaces to use a
separable collective Hamiltonian such as $H_\text{coll}$, which is designed for
complete oscillator shells.  In the absence of a good collective interaction, we
proceed along two rather extreme paths, with the idea that the size of isoscalar
pairing effects will be somewhere between what the two paths yield.  Our first
procedure is to set all $J^\pi=1^+,T=0$ two-body matrix elements equal to zero
in our shell model Hamiltonian.
Because these interaction matrix elements receive contributions not only
from isoscalar pairing but also from other modes, removing them from the shell-model
Hamiltonian may overestimate the effects of isoscalar pairing.
Our second procedure is to subtract from the shell model interaction the isoscalar
pairing interaction from $H_\text{coll}$,
even though the incomplete oscillator shells limit its effectiveness.
Because of the limitation, the resulting interaction probably underestimates
the effects of isoscalar pairing. Following the recommendation of 
Refs.~\cite{duf96,ber10}, we use the same isoscalar pairing strength as in 
Table~\ref{tab:couplings} in these heavier nuclei. 

\begin{figure}
\includegraphics[width=\columnwidth]{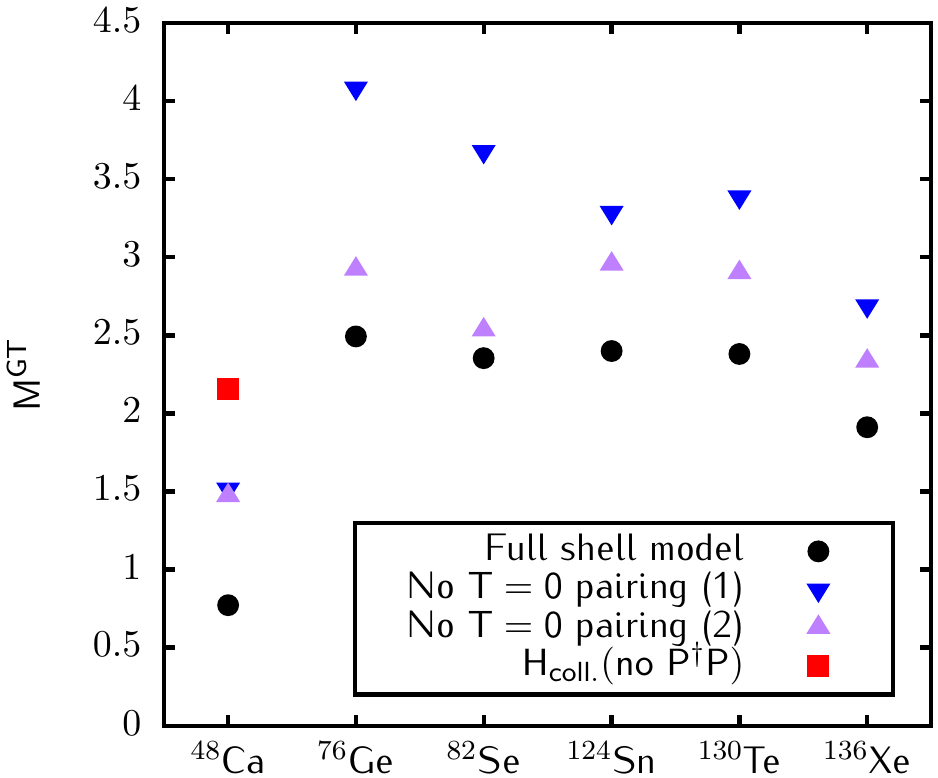}
\caption{Estimates of the effect of isoscalar pairing on \bbz
decay for nuclei used in or considered for experiments.  The Gamow-Teller matrix
elements, $M_{0\nu}^{GT}$, are shown for the full shell model effective
interaction (black circles), the effective interaction with all $J^\pi=1^+,T=0$
two-body matrix elements set to zero (blue upside-down triangles), and the
effective interaction with the isoscalar pairing interaction from
$H_\text{coll}$ subtracted (purple triangles).  For $^{48}$Ca the result from
Fig.~\ref{fig:ca_pqq}, obtained from $H_\text{coll}$ without the isoscalar
pairing term, is also shown (red square).
}\label{fig:real} 
\end{figure}

The results appear in Fig.~\ref{fig:real}.  In $^{48}$Ca, for which we include
all spin-orbit partners, the two prescriptions give very similar matrix
elements, larger by about a factor of two than the KB3G matrix element, but
smaller than the matrix element obtained with $H_\text{coll}$ and excluding
isoscalar pairing.  This last result
suggests (again) that additional perhaps non-collective correlations, present in
the full shell-model interaction but not included in $H_\text{coll}$, may in
part make up for the removal of isoscalar pairing correlations.

For heavier isotopes the two prescriptions produce different results.  Setting
all $J^{\pi}=1^+, T=0$ interaction matrix elements equal to zero increases the
\bbz decay matrix elements by 1 to 2 units, or between 60\% and 80\%.  The effect is
smaller in nuclei with larger isospin.  On the other hand, subtracting the
isoscalar pairing interaction in $H_\text{coll}$ from the shell-model
interactions leads to much smaller increases, of 25\% or less.
We conclude that isoscalar pairing in heavier \bb decay
candidates is important, but probably less so than in the
lighter isotopes studied in Sec.\ \ref{sm_results}, as suggested by the 
range covered by the two prescriptions considered. This conclusion is
tentative, however, because the effects of single-particle
orbitals beyond the valence space cannot be completely
assessed without including them explicitly.

\section{Summary}~\label{summary}

We have explored the role of collective correlations in \bbz decay.  To be able
to study all relevant collective terms within the shell model we have mostly
limited ourselves to nuclei in the $pf$-shell, with $A=60$ at most.  We have
found that a separable collective Hamiltonian that includes a monopole term,
isovector and isoscalar pairing, a quadrupole-quadrupole interaction and a
spin-isospin term reproduces the matrix elements obtained with the full shell
model interaction quite well.  Among the collective terms, isoscalar pairing has
a particularly strong effect, one that can be related to the approximate $SU(4)$
symmetry obeyed by the \bbz decay operator and the nuclear interaction.  In heavier
nuclei, the effects of isoscalar pairing are harder to estimate and probably
smaller than in the $pf$-shell, but almost certainly still important.

GCM calculations with the same collective Hamiltonian for $pf$-shell nuclei
agree well with full shell model calculations provided that the $T=0$ pairing
amplitude is one of the generator coordinates.  The agreement suggests that
theories of collective motion such as the GCM with a few generator coordinates
may provide accurate matrix elements in heavier nuclei.  If collective
correlations from beyond the valence space turn out to be important, such models
will be particularly useful because shell-model calculations in several
oscillator shells are still computationally prohibitive. 

\section*{Acknowledgements}

We thank A. Poves and A. Schwenk for useful discussions.
This work was supported in part by an International Research Fellowship from
the Japan Society for the Progress of Science (JSPS), and JSPS KAKENHI grant
No. $26\cdot04323$, by the Deutsche Forschungsgemeinschaft through contract
SFB~634, by the Helmholtz Association through the Helmholtz Alliance Program,
contract HA216/EMMI ``Extremes of Density and Temperature: Cosmic Matter in the
Laboratory'', by the European Research Council under grant 307986 STRONGINT,
by the U.S.\ Department of Energy through Contract No.  DE-FG02-97ER41019,
and by the Spanish MINECO under Programa Ram\'on y Cajal 11420 and FIS-2014-53434-P.
Numerical calculations were performed in part on the COMA (PACS-IX) System at
the Center for Computational Sciences, University of Tsukuba.

%

\end{document}